\documentclass[aps,pre,twocolumn,showpacs,superscriptaddress,
amsmath,amssymb]{revtex4}
\usepackage{graphicx}
\usepackage{dcolumn}
\usepackage{bm}
\begin{document}
\title{Surface and capillary transitions in an associating binary mixture 
model.}   
\author{J. M. Romero-Enrique}
\email[Corresponding author: ]{jose.enrique@ic.ac.uk}
\affiliation{Department of Mathematics, Imperial College of Science, Technology
and Medicine, 180 Queen's Gate, London SW7 2BZ, United Kingdom}
\affiliation{Departamento de F\'{\i}sica At\'omica, Molecular y Nuclear, 
Area de F\'{\i}sica Te\'orica, Universidad de Sevilla, Apartado de Correos 
1065, 41080 Sevilla, Spain}
\author{L. F. Rull}
\affiliation{Departamento de F\'{\i}sica At\'omica, Molecular y Nuclear, 
Area de F\'{\i}sica Te\'orica, Universidad de Sevilla, Apartado de Correos 
1065, 41080 Sevilla, Spain}
\author{U. Marini Bettolo Marconi}
\affiliation{Dipartimento di Fisica and Istituto Nazionale di 
Fisica della Materia, via Madonna delle Carceri, Universit\`a di Camerino, 
62032 Camerino, Italy}
\begin {abstract}
We investigate the phase diagram of a two-component associating 
fluid mixture in the presence of selectively adsorbing substrates. 
The mixture is characterized by a bulk phase diagram which displays peculiar
features such as closed loops of immiscibility. The presence of the substrates
may interfere the physical mechanism involved in the appearance of these phase
diagrams, leading to an enhanced tendency to phase separate below the lower
critical solution point. Three different cases are considered: a planar 
solid surface in contact with a bulk fluid, while the other two represent 
two models of porous systems, namely a slit and an array on infinitely long 
parallel cylinders. We confirm that surface transitions, as well as 
capillary transitions for a large area/volume ratio, are stabilized in 
the one-phase region. Applicability of our results to experiments reported
in the literature is discussed.   
\end{abstract}
\pacs{64.75.+g, 05.70.Np, 68.08.Bc, 68.35.Rh}
\maketitle
\section{Introduction}
In the past few years there has been a considerable progress
in the understanding of the physics of fluids in restricted geometries 
\cite{Evans}. The properties of liquids or gases confined in narrow pores
appear different from those in the bulk. A common trend of these systems
is the fact that condensation occurs at pressures different from the 
saturation bulk pressure and the critical temperature for phase separation 
is generally lower than the corresponding bulk temperature. According to 
Nakanishi and Fisher \cite{Nakanishi1,Nakanishi2,Nakanishi3,Nakanishi4} an 
incompressible fluid mixture (with isotropic interactions) under the effect 
of confinement displays a reduced tendency toward phase separation at a 
fixed temperature, because the effective attractive forces result weaker 
within a pore. The confinement also determines an effective dimensional 
reduction thus enhancing fluctuation effects. In addition, the presence of 
adsorbing walls may induce a surface transition in a sub-critical fluid, 
even when this is in a single phase region in the bulk. Such a behavior, 
which is now well understood by means of a mean field approach 
\cite{EUT,Ev1,Ev2} is due essentially to the reduction of the contribution 
to the free energy from the the cohesive fluid-fluid forces caused by the 
presence of the confining walls. There exists, however, a class of 
non-simple fluid mixtures for which the situation is different: these are 
the strongly associating mixtures. Strong directional attractive interactions 
such as hydrogen bonding or charge transfer complexing between molecules 
affect dramatically the properties of the fluid or solid.
An interesting phenomenon that occurs in such a mixture is the reentrant 
miscibility: the mixture is completely miscible at low temperatures, up to
a temperature where it separates into two liquid phases. As the temperature
is further increased the complete miscibility reappears. The nicotine-water
mixture is a paradigm for such a behavior \cite{Hudson}.
The temperature-composition phase diagram shows closed-loops of 
immiscibility, characterized by the existence of a lower and an 
upper critical solution point (LCSP and UCSP, respectively). 
Hirschfelder \emph{et al} \cite{Hirschfelder} proposed a mechanism to explain
the reentrant miscibility. The essential ingredient is the existence 
of highly directional strong attraction (as the hydrogen-bond) between 
the unlike particles. The dispersive (isotropic) forces between these 
particles are assumed to be so weak as to favor segregation of the species.  
At low temperatures, the energy favors complete mixing. However, as the 
temperature increases, the energy decrement due to the formation of 
directional bonds competes with the decrease of orientational entropy.
As a consequence, the system phase separates when the temperature raises.
This mechanism has been confirmed by theoretical studies on lattice
\cite{Barker,Wheeler,Walker} and continuous models \cite{Jackson}, as 
well as in recent computer simulation studies \cite{Davies1,Davies2}.

How the presence of substrates or the confinement in a pore
affect the behavior of these mixtures (if in a manner similar to that of 
simple fluids, or if on the contrary these mixtures are peculiar) is an 
issue of great practical and theoretical interest.
The presence of selectively adsorbing substrates can interfere with the 
formation of bulk hydrogen-bonding, leading to a stabilization 
of the two-phase region with respect to the completely miscible phase. 
Computer simulation studies on confined lattice models
\cite{Kumar} show the appearance of surface transitions in a range of 
temperatures below the LCSP. The same mechanism has been invoked 
to explain some experiments \cite{Scheibner,Meadows} that seemed to 
contradict the Nakanishi-Fisher picture. 

In the present paper we shall study a lattice model of associating binary 
mixtures, introduced few years ago by Lin and Taylor \cite{Lin1,Lin2} and 
reconsidered recently by the present authors \cite{comment,Romero,
Romero2,Romero3}. In spite of its idealized nature, the model has the 
merit of allowing for an explicit analysis of its non trivial phase diagram. 
As an instance, the study of its bulk properties predicts the existence of 
closed-loops of immiscibility under constant pressure conditions. 
This analysis will be extended to the behavior of the mixture in presence of
substrates and under confinement. 

The structure of the paper is organized as follows. In section II we 
briefly present the model and write explicitly the equations for the 
order parameter profiles. Section III is devoted to analyze in detail the 
wetting behavior of the mixture in the presence of adsorbing substrates. 
In section IV we study the capillary behavior of the same system when 
confined between two parallel plates (a slit) and in section V we 
consider the same system confined in a network of very narrow cylinders. 
Capillary transitions (and surface transitions for the slab geometry) will 
be analyzed in terms of the physical parameters of the system. Finally, in 
section VI we present our conclusions.

\section{The model}

\begin{figure}
\includegraphics[width=8.6cm]{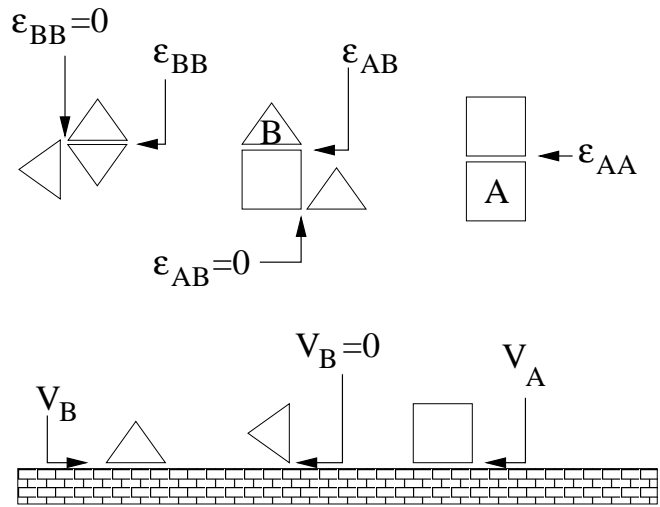}
\caption{Schematic representation of the A-B LT model in the 
two-dimensional square lattice (the underlying lattice is not shown 
for clarity). Relevant fluid-fluid and fluid-substrate interactions 
are also highlighted (see text for explanation). \label{fig1}}
\end{figure}

Some years ago, Lin and Taylor proposed a lattice model to describe a 
binary mixture in which the two unlike species can form highly directional 
bonds. In more detail, the Lin-Taylor (LT) model \cite{Lin1,Lin2} is an 
A-B binary mixture lattice model, defined on a generic D-dimensional lattice
of coordination number $\nu$. The A particles are allowed to singly occupy 
any of the lattice cells. Every cell contains $\nu$ equal sub-cells, one 
for each face of the mother cell. Again, each sub-cell can host at most 
one B particle, provided the mother cell is not occupied by an A particle. 
In addition to the repulsive interactions (which determine the previous 
occupation rules), one includes short-ranged interactions 
(see Fig. \ref{fig1}). The interaction strength between a pair of 
nearest-neighbor A particles is $\epsilon_{AA}$, $\epsilon_{AB}$ between 
unlike species and $\epsilon_{BB}$ between B-B particles.
All these interactions are non-vanishing only if the particles
share a face.

The bulk properties, as well as the global phase diagram, have been 
exhaustively analyzed in previous work \cite{comment,Romero,Romero2,Romero3}. 
One of the main conclusions of such studies is that the appearance 
of closed loops of immiscibility occurs when the interaction between 
A-A and A-B pairs are attractive ($\epsilon_{AA}<0$, $\epsilon_{AB}<0$), 
and $\epsilon_{AB}\lesssim (\epsilon_{AA}+\min(\epsilon_{BB},0))/2$. 
The latter condition is in complete agreement with the mechanism proposed 
by Hirschfelder \emph{et al} \cite{Hirschfelder} for the reentrant miscibility.
In this work, we assume the bulk fluid-fluid parameters to have values: 
$\epsilon_{AA}=-1$ (it defines the energy scale), $\epsilon_{AB}=-0.65$ 
and $\epsilon_{BB}=0$, so that we are in the regime where closed-loops 
of immiscibility appear.  

The presence of substrates introduces new interactions (see Fig.
\ref{fig1}). We shall assume that the substrates do not 
affect the fluid-fluid interactions, i.e. there is no surface enhancement. 
The interaction strength between an A particle and the substrate is $V_A$,
if the particle is in a cell in contact with the substrate, and zero
otherwise. The interaction between the B particles and the substrate is 
orientationally dependent, i.e. its strength $V_B$ is non vanishing only 
if the face of the B particle touches the substrate. The form of the 
substrate-fluid interactions are consistent with the bulk interactions. 
Hereafter, the substrate-fluid interactions will be assumed as
attractive, i.e. $V_A\le 0$ and $V_B\le 0$. For absolute 
values of $V_B$ large enough, the substrate-B interactions compete with 
the bulk A-B interactions, so that one expect this type of 
substrate to enhance the demixing tendency for temperatures lower 
than the corresponding LCSP. 

To study the LT model in presence of substrates, we consider the 
grand-canonical ensemble, where the system is in thermal and chemical 
equilibrium with a reservoir at fixed temperature $T$ and chemical potentials
$\mu_A$ and $\mu_B$. The activities $z_A$ and $z_B$ are defined via 
$z_A\equiv \exp(\beta \mu_A)$ and $z_B\equiv \exp(\beta \mu_B)$, where
$\beta=1/k_B T$. After tracing over all configurations of the B particles, 
the system results to be isomorphic to a mono-component lattice gas, with 
renormalized interactions and chemical potential. The details on this procedure
can be found in Refs. \cite{Lin2,Romero,Romero3}. The grand-canonical 
partition function $\Xi$ of the LT model can be written in terms of an 
equivalent Ising model as:
\begin{eqnarray}
\Xi &=& \left(1+2z_B+z_B^2 {\rm e}^{-\beta \epsilon_{BB}}
\right)^{\frac{\nu N}{2}}\exp \left( N \left(H - \frac{\nu K}{2}\right)\right)
\nonumber\\ 
&\times&
\left(\frac{1+z_B \mathrm{e}^{-\beta V_{B}}}
{\sqrt{1+2z_B+z_B^2 \mathrm{e}^{-\beta \epsilon_{BB}}}}
\right)^{\nu_\perp N_b} \exp \Bigg[N_b\Bigg(\Delta H_1 
\nonumber\\ 
&+&\frac{\nu_\perp K}{2}\Bigg)
\Bigg]
Z_{D,\nu}^{Ising}(K,H,\Delta H_1)
\label{partition}
\end{eqnarray}
where $N$ is the total number of lattice cells, $N_b$ the number of 
cells adjacent to the substrate, $\nu_\perp$ the number of nearest-neighbors  
in the direction perpendicular to the substrate; $\nu_\parallel$ is
defined as the number of nearest neighbor cells in directions parallel to the
substrate. Obviously, $\nu=\nu_\parallel+2\nu_\perp$. 
Finally, $Z_{D,\nu}^{Ising}(K,H,\Delta H_1)$ is the canonical partition 
function of the Ising model on the same lattice. Let us name 
$K$ the effective coupling constant, $H$ the effective bulk magnetic field, 
and $\Delta H_1$,  the effective surface magnetic field 
acting only on the boundary sites:
\begin{eqnarray}
K&=& - \frac{\beta \epsilon_{AA}}{4}+
\frac{1}{2} \ln \left( \frac{\sqrt{1 + 2z_B+z_B^2 {\rm e}^{-\beta
\epsilon_{BB}}}}{1+z_B \textrm{e}^{-\beta
\epsilon_{AB}}}\right) \label{defk} 
\\
\nonumber\\
H&=& \frac{1}{2} \ln z_A- \frac{\nu \beta \epsilon_{AA}}{4}
-\frac{\nu}{4} \ln (1+2z_B+z_B^2 \mathrm{e}^{-\beta \epsilon_{BB}}) 
\label{defh}
\nonumber\\
\\
\Delta H_1&=&\nu_\perp\Bigg[
\frac{1}{2} \ln \left(\frac{\sqrt{1+2z_B+z_B^2 \mathrm{e}
^{-\beta \epsilon_{BB}}}}{ 1+z_B \mathrm{e}^{-\beta V_B}}\right)
\nonumber\\
&+&\frac{\beta}{4}\left(\epsilon_{AA}-\frac{2V_A}{\nu_\perp}\right) 
\Bigg]
\label{H_1}
\end{eqnarray}
All equilibrium properties can be obtained from the knowledge of the 
partition function Eq. (\ref{partition}) by using standard thermodynamical 
relationships. In particular, the bulk pressure is obtained as:
\begin{eqnarray}
p&=&\lim_{V\to \infty}\! k_B T \frac{\ln \Xi}{V}=
\frac{k_B T}{v_0} \Bigg[\frac{\nu}{2}\ln \left(1+2z_B+z_B^2 \mathrm{e}
^{-\beta \epsilon_{BB}}\right)\nonumber\\
&-&\frac{\nu K}{2}+H+\lim_{N\to \infty} \frac{1}
{N}\ln Z_{D,\nu}^{Ising}\Bigg]
\label{pressure}
\end{eqnarray}
where $v_0$ is the volume of lattice cell, and $N\equiv V/v_0$.
The mole fraction $X_A(i)$ profile is obtained as:
\begin{equation}
X_A(i)\equiv \frac{n_A(i)}{n_A(i)+\sum_{s=1}^\nu n_B(i,s)}
\label{fracprof}
\end{equation}
where $n_A(i)$, the A particle density in the cell $i$, is defined in terms
of the local magnetization $m_i$ as:
\begin{equation}
n_A(i)=\frac{1+m_i}{2v_0}
\label{naprof}
\end{equation}
and $n_B(i,s)$, the B particle density in the sub-cell $s$ of the cell $i$ is
\begin{equation}
n_B(i,s)=\frac{1-m_i}{2v_0}\frac{z_B \mathrm{e}^{-\beta V_B}}
{1+z_B \mathrm{e}^{-\beta V_B}}
\label{nbprof1}
\end{equation}
if the sub-cell is by the substrate (see Fig. \ref{fig1}). Otherwise, 
$n_B(i,s)$ is defined as:
\begin{eqnarray}
n_B(i,s) = \frac{1}{4 v_0} \Bigg[ \frac{z_B {\rm e}^{-\beta \epsilon_{AB}}
}{1 + z_B \mathrm{e}^{-\beta \epsilon_{AB}}} (1 - m_i &+& m_j - 
\langle s_i s_j
\rangle) \nonumber \\
+ \frac{z_B + z_B^2 \mathrm{e}^{-\beta \epsilon_{BB}}}{
1 + 2z_B + z_B^2 \mathrm{e}^{-\beta \epsilon_{BB}}} (1 - m_i - m_j &+& 
\langle s_i s_j\rangle) \Bigg]
\label{nbprof2}
\end{eqnarray}
where $\langle s_i s_j\rangle$ is the 2-spin correlation function between the
magnetization on the cell $i$ and its nearest-neighbor $j$ that it is
associated to the sub-cell $s$. It is easy to see that in the homogeneous
case these equations reduce to the expressions given in Refs. 
\cite{Romero,Romero3}. 

The equivalence between the LT model and the Ising model, allows one to 
obtain the surface behavior of the mixture from the knowledge of the behavior
of the latter. However, such a mapping, is not at all trivial and may 
lead in some cases to unexpected results.

\section{Wetting transitions in the LT model}

Let us consider an A-B binary mixture in the presence of a planar, infinite 
and structureless substrate. Such a substrate confines the fluid to the 
half-space $z \ge 0$. When the fluid (e.g., the A-particle rich phase,
$\alpha$) is at bulk coexistence, the substrate can promote the formation 
of drops of the opposite coexisting phase (the B-particle rich phase $\beta$) 
on it, characterized by a contact angle $\theta$. When $\theta 
=0$, the drops spread over the substrate, i.e. the phase $\beta$ wets 
completely the $\alpha$-substrate interface. A finite value of $\theta$, 
instead, corresponds to a partial wetting situation. 
The crossover between a partial to a complete wetting regime is called
the wetting transition \cite{Dietrich}. The surface tensions involved in the 
problem ($\sigma_{\alpha s}$, $\sigma_{\beta s}$ and $\sigma_{\alpha \beta}$,
corresponding to the $\alpha$-substrate, $\beta$-substrate and
$\alpha$-$\beta$ interfaces, respectively) are related to the contact angle
via the Young equation \cite{Widom}:
\begin{equation}
\sigma_{\alpha s}=\sigma_{\beta s} + \sigma_{\alpha \beta} \cos \theta
\label{Young}
\end{equation}
Thermodynamically the surface tensions correspond to the excess
grand-canonical free energy per interfacial unit area (respect to the bulk)
\cite{Widom}. Since both $\sigma_{\alpha \beta}$ and $\sigma_{\beta s}$ have an
analytical behavior at this thermodynamic state,
the wetting transition is a (surface) thermodynamic
phase transition, as it can be seen from Eq. (\ref{Young}) and the dependence
of $\theta$ on the temperature. As usual, appropiate
order parameters corresponding to the wetting transition can be defined as
first derivatives of the surface tension respect to thermodynamic fields.
In simple fluids, the excess number of particles per interfacial unit area or
adsorption $\Gamma$ (that corresponds to $-(\partial \sigma_{\alpha s}/
\partial \mu)_T$, being $\mu$ the chemical potential) is the natural order
parameter choice. In mixtures, other choices are available, e.g. the 
volume integrated excess mole fraction per interfacial unit area. 
Microscopically, 
the wetting transition manifests itself as the growth of a macroscopically 
thick $\beta$ layer intruding in the $\alpha$-substrate interface 
composition profile. 
Three different types of wetting transition may arise. If the layer width 
jumps suddenly from a microscopically finite value, the wetting transition 
is first order, and an off-coexistence surface transition (prewetting) is 
associated to it. If the layer width diverges continuously, the wetting 
transition is named critical. Finally, the wetting transition can occur as 
the accumulation point of an infinite sequence of first-order transitions 
in the layering wetting case (these transitions extend to the 
off-coexistence region).
The conditions for such a transition depend subtly 
on the fluid-fluid and substrate-fluid interactions, and interfacial 
fluctuations can play an important role \cite{Lipowsky}.  

\begin{figure}
\includegraphics[width=8.6cm]{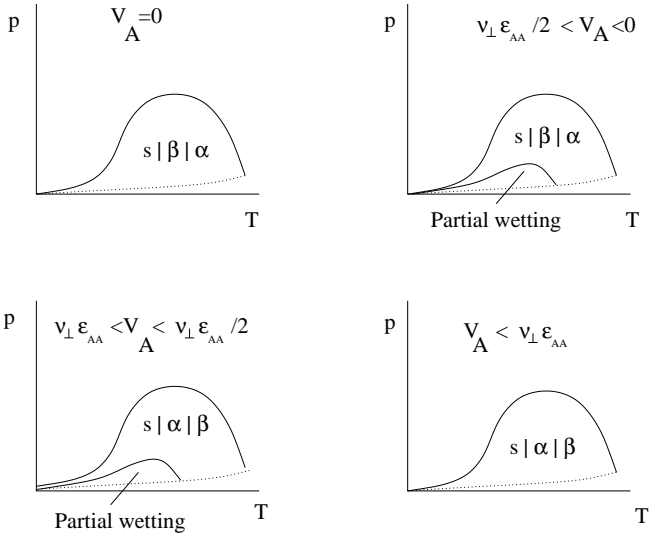}
\caption{Qualitative wetting phase diagrams for the LT model for $V_B=0$.
Thick solid lines correspond to critical lines, dashed lines to the 
A pure gas-liquid coexistence lines and thin solid lines to the wetting 
transition lines. The symbol $s|\alpha|\beta$ ($s|\beta|\alpha$) means 
the bulk conditions where there exist wetting of the $\beta$-substrate 
($\alpha$-substrate) interface by the $\alpha$ ($\beta$) phase, respectively 
(see text). The bulk conditions under which there is partial wetting are 
also shown.\label{fig2}}
\end{figure}

In the LT model two-phase coexistence occurs when $H=0$ and $K>K_c$, where
$K_c$ is the lattice dependent Ising critical coupling. Moreover, the 
wetting behavior is controlled by the ratio $\Delta H_1/\nu_\perp K$, 
that is the ratio between the surface energy and the loss of bulk energy 
due to the presence of the surface. If such a parameter is positive the 
substrate favors the $\alpha$-phase to intrude between the substrate and 
the $\beta$-phase. On the other hand, if $\Delta H_1/\nu_\perp K<0$ the 
phase $\beta$ intrudes between the substrate and the phase $\alpha$. 
We shall focus on the coexistence side $H=0^+$ ($H=0^-$) for 
$\Delta H_1 < 0$ ($\Delta H_1> 0$), respectively (otherwise, only 
partial wetting can occur). If the absolute value of 
the parameter is larger than one ($|\Delta H_1/\nu_\perp K|>1$),
then all coexistence surface corresponds to a complete wetting situation. 
If such a condition is not fulfilled, one will observe
a partial wetting situation up to a value $K_w$ (dependent on 
$|\Delta H_1/\nu_\perp K|$) and beyond it a complete 
wetting situation up to $K=K_c$.   

Let us consider first the case $V_B=0$, when $\Delta H_1$
depends only on $T$, regardless the value of $z_B$. Moreover,
the sign of $\Delta H_1$ is uniquely determined by the difference between 
$V_A$ and  $\nu_\perp \epsilon_{AA}/2$ over the whole temperature range. 
One can envisage various cases (see Fig. \ref{fig2}): if $V_A=0$, 
the coexistence corresponds to a situation of complete wetting of the 
interface substrate-$\alpha$ phase by phase $\beta$, denoted by the 
symbol $s|\beta|\alpha$. If $\nu_\perp\epsilon_{AA}/2 < V_A < 0$, there 
exists a region of complete wetting $s|\beta|\alpha$ and a lower 
pressure region of partial wetting. As $V_A$ decreases, the region of 
complete wetting $s|\beta|\alpha$ shrinks and eventually disappears 
for $V_A=\nu_\perp\epsilon_{AA}/2$, because $\Delta H_1 \equiv 0$. 
For $V_A < \nu_\perp\epsilon_{AA}/2$, the sign of $\Delta H_1$ changes 
and a region of complete wetting of the interface substrate-phase 
$\beta$ by phase $\alpha$ appears ($s|\alpha|\beta$). As $V_A$ increases, 
the region of partial wetting reduces and disappears for 
$V_A \ge \nu_\perp \epsilon_{AA}$. 

\begin{figure}
\includegraphics[width=8.6cm]{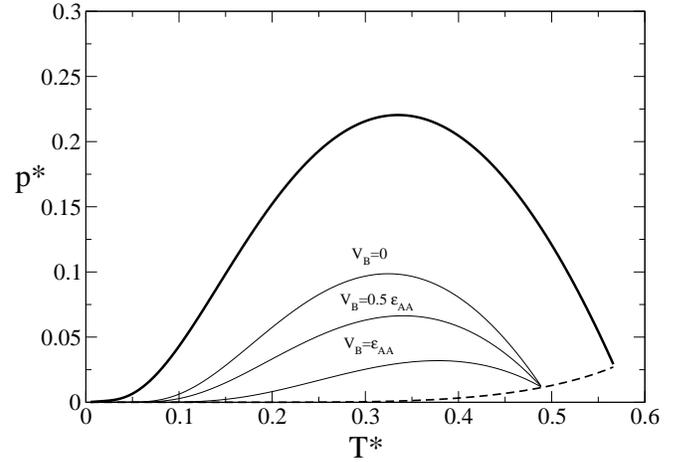}
\caption{Wetting phase diagram for the square lattice LT model. The 
bulk parameters are fixed to $\epsilon_{AB}=0.65 \epsilon_{AA}<0$, 
$\epsilon_{BB}=0$, and $V_A=0.25\epsilon_{AA}$. The thick solid 
line corresponds to the bulk critical line, and the thick dashed line to 
the liquid-vapor transition line of the A pure fluid. The thin solid 
lines correspond to the wetting transition lines for different values of $V_B$.
Hereafter the reduced units are defined via $T^*\equiv k_B T / 
|\epsilon_{AA}|$ and $p^*\equiv p v_0 / |\epsilon_{AA}|$, where $v_0$ 
is the volume of a lattice cell. See explanation in text. \label{fig3}}
\end{figure}

When $V_B\ne 0$, it is possible that in correspondence of different 
state points the substrate favors  either the $\beta$ or the $\alpha$ phase.
To study such an effect it is necessary to locate the coexistence states 
where $\Delta H_1=0$. Since $V_B$ and $\epsilon_{AB}$ are negative, 
for a temperature $T$ all the coexistence states with pressure $p$ 
larger than $p_{nw}(T)\equiv p(T,H=0,\Delta H_1=0)$ correspond to 
$\Delta H_1 > 0$, whereas states with lower pressures correspond to 
$\Delta H_1 <0$. It is easy to prove from  Eq. (\ref{H_1}) that a 
necessary condition for the existence of this line is that 
$V_B+\epsilon_{AA}/2 < (V_A/\nu_\perp) \le \epsilon_{AA}/2$. However, 
$\Delta H_1$ changes sign for a temperature $T$ only if the value of 
the pressure which corresponds to $\Delta H_1 = 0$ is less than the 
critical pressure at such temperature. 

In order to illustrate such a phenomenology, we shall study the 2D square 
lattice case ($\nu=4$, $\nu_\perp=1$). In this case, only critical wetting 
can occur (fluctuation effects preclude any other possibility) and the 
value of $K_w$ at which the transition of critical wetting occurs is given 
by solution of the equation
\cite{Abraham}
\begin{equation}
\exp(2K_w)[\cosh(2K_w)-\cosh(2\Delta H_1)]=\sinh(2K_w)
\label{H1wet}
\end{equation}
Let us first consider the case $\nu_{\perp}\epsilon_{AA}/2 \le V_{A} < 0$: 
$\Delta H_1<0$, for arbitrary values of $T$, $z_B$ y $V_B$.  
Moreover, $\Delta 
H_1$ becomes more negative as $V_B$ decreases, at fixed $T$ and $z_B$ 
(or pressure $p$). The typical diagrams of complete wetting are displayed 
in Fig. \ref{fig3}. The region between the wetting 
transition line and the critical line corresponds to a situation of 
wetting $s|\beta|\alpha$, and the remaining coexistence region 
corresponds to a situation of partial wetting. The complete wetting zone 
increases as $V_B$ becomes more negative. 

\begin{figure}
\includegraphics[width=8.6cm]{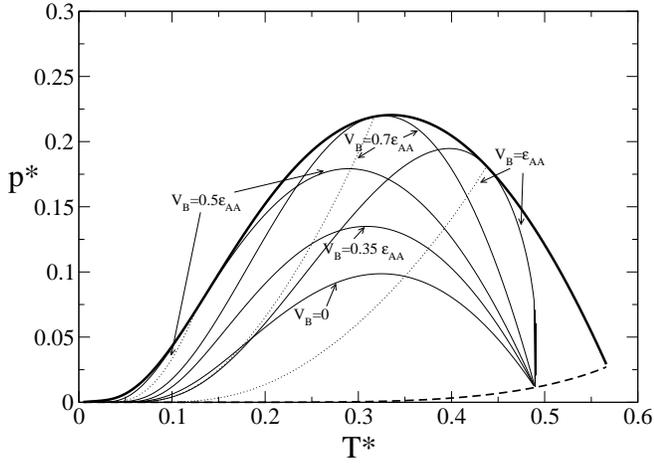}
\caption{The same as Fig. \ref{fig3} but for $V_A=0.75 \epsilon_{AA}$. 
The dotted lines correspond to the coexistence points at which 
$\Delta H_1=0$. See text for explanation.\label{fig4}}
\end{figure}

\begin{figure}
\includegraphics[width=8.6cm]{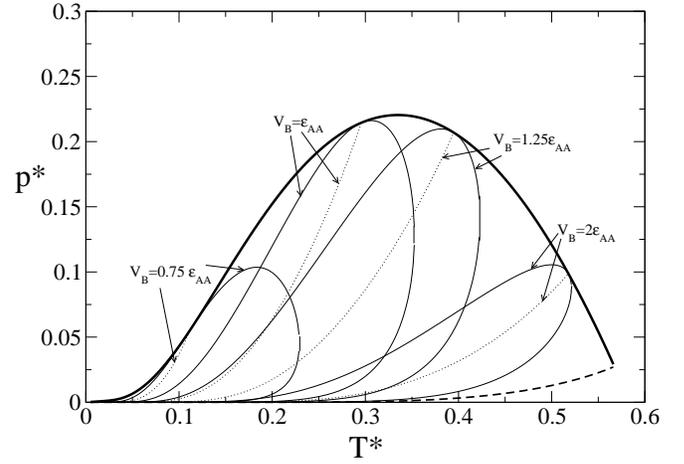}
\caption{The same as Fig. \ref{fig4} but for $V_A=\epsilon_{AA}$. See text
for explanation.\label{fig5}}
\end{figure}

When $(V_A/\nu_\perp)<\epsilon_{AA}/2$, the substrate also favors
the adsorption of particles A. This determines a competition between 
A and B species. In general, when $(V_A/\nu_\perp)-\epsilon_{AA}/2<
0.1\epsilon_{AA}$ $\beta$-phase preferential adsorption can only occur if
$V_B-(V_A/\nu_\perp)<\epsilon_{AB}-\epsilon_{AA}$. Figures \ref{fig4} 
and \ref{fig5} display the wetting diagrams for 
$\epsilon_{AA}<(V_A/\nu_\perp)<\epsilon_{AA}/2$ and $(V_A/\nu_\perp)
\le \epsilon_{AA}$, respectively. In the first case, the complete 
wetting transition line originates at $T=0$ and terminates at the 
temperature of complete wetting for the A-pure liquid in the interface 
substrate-A vapor. For values of $V_B \simeq 0$, $\Delta H_1 > 0$ for 
all conditions of coexistence, and the configurations are $s|\alpha|\beta$. 
For $V_B<(V_A/\nu_\perp)+\epsilon_{AB}-\epsilon_{AA}$ there appear 
states characterized by $\Delta H_1<0$. As a consequence, within 
the coexistence region at the left of the line $p_{nw}(T)$ 
the substrate adsorbs preferentially particles B, and the 
wetting is of type $s|\beta|\alpha$. On the other hand,
at the right of the curve $p_{nw}(T)$ the wetting is of type
$s|\alpha|\beta$. The line of wetting transition touches tangentially the 
critical curve at the point where the curve in turn $p_{nw}(T)$ crosses the
critical line. Such a point moves monotonically toward higher temperatures
as $V_B$ decreases, driving with it the wetting transition line and the
$p_{nw}(T)$ curve. As also happened for the $\nu_\perp \epsilon_{AA}/2 < 
V_A <0$ case, the wetting transition line from a $s|\beta|\alpha$ complete
situation to partial wetting moves to lower pressures, converging toward
the A-pure liquid-vapor coexistence line for $V_B\to -\infty$ (however, along
this line there is complete wetting by liquid of the vapor-substrate interface
over all the temperature range).

When $V_A \le \nu_\perp \epsilon_{AA}$ (see Fig. \ref{fig5}), for values 
$V_B$ near zero, the coexistence corresponds to complete wetting 
$s|\alpha|\beta$. Only when $V_B< (V_A/\nu_\perp)+\epsilon_{AB}-
\epsilon_{AA}$ states of partial wetting reappear, near the line
$p_{nw}(T)$. The behavior of the latter as well as the wetting transition
line is similar to the one observed in the previous case, with the 
difference that the right branch of the wetting transition line (that 
corresponds to a transition from partial to $s|\alpha|\beta$ complete wetting)
begins at zero temperature. 

One expects that the behavior of the wetting transition for the 
three-dimensional case will be qualitatively similar to the two-dimensional 
case. However, the complete wetting phase diagram for the three-dimensional
Ising model is not totally understood yet. Theoretical \cite{Pandit}
and simulation studies \cite{Binder} show, in the case of strictly 
short-ranged forces and in the absence of an enhanced 
surface spin-spin coupling, that the wetting transition is continuous 
only if $K_w$ is smaller than $K_R$, where $K_R$ is the value of the 
coupling constant corresponding to the roughening transition ($K_R\approx 
1.85 K_c$ for the simple cubic lattice). However, if such condition is not 
fulfilled, the wetting transition will occur through a sequence of first 
order layering transitions. Theoretical studies have claimed that the 
wetting transition can be weakly first-order instead of second order 
\cite{Jin-Fisher}. In the present work we will consider the Ising model 
in the framework of the mean-field Bragg-Williams approximation:
\begin{eqnarray}
& &\lim_{N_\perp \to \infty}
\frac{1}{N_\perp}\ln Z_{Ising}(K,H,\Delta H_1)= \Delta H_1 m_1 
\nonumber\\
& & 
+\sum_{j=1}^{\infty}\Bigg[\frac{\nu_\parallel}{2} K m_j^2
+ \nu_\perp K m_j m_{j+1} +H m_j \nonumber\\
& &
-\left(\frac{1+m_j}{2}\right) \ln \left(\frac{1+m_j}{2}\right)
\nonumber\\
& &
- \left(\frac{1-m_j}{2}\right) \ln \left(\frac{1-m_j}{2}\right)\Bigg] 
\label{meanfield}
\end{eqnarray}
where $N_\perp$ is the number of cells in a layer, an $\{m_j\}$, $j=1,2,
\ldots\infty$ is the equilibrium magnetization profile obtained by 
solving the Euler-Lagrange equations:
\begin{equation}
m_j = \tanh(\nu_\parallel K m_j + \nu_\perp K (m_{j-1}+m_{j+1}) 
+ H)\  \label{E-L}
\end{equation}
where $m_0\equiv \Delta H_1/\nu_\perp K$. This set of equations must be solved
in conjunction with the asymptotic condition $m_j \to m_b$ for $j \to 
\infty$, where $m_b$ is the spontaneous magnetization per site
in the bulk case with coupling constant $K$ and magnetic field $H$. 
In general, different solutions of Eq. (\ref{E-L}) are found. The equilibrium
profile is the solution that maximizes the functional (\ref{meanfield}).
This approach neglects capillary fluctuations, that it is equivalent 
to assume $K_R \equiv K_c$ and the wetting transition occurs as a sequence
of layering transitions \cite{Oliveira,Ebner}. Once the equilibrium 
magnetization profile is computed, the surface tension $\sigma$ is 
obtained via:
\begin{equation}
\sigma=(\Omega + pV)/A
\label{tens-LT}
\end{equation}
where $\Omega=-k_B T \ln \Xi$, with $\Xi$ defined by the 
Eqs. (\ref{partition}) and (\ref{meanfield}), and the pressure $p$ by 
Eq. (\ref{pressure}). The adsorption $\Gamma$, that it is the relevant 
order parameter in the wetting transition, is defined as:
\begin{equation}
\Gamma = \sum_{j=1}^\infty (X_A(j)-X_{A}^b)
\label{adsorcion}
\end{equation}
where $X_A(j)$ is the A molar fraction profile Eq. (\ref{fracprof}) by using
the approximation $\langle s_i s_j\rangle \approx m_i m_j$. Finally,  
$X_{A}^b$ is its bulk value.

We have studied the simple cubic lattice, for which $\nu=6$ and 
$\nu_\perp=1$. The magnetization profiles have been obtained by an
iterative method of the Eq. (\ref{E-L}) for $j=1,\ldots,N_{max}$, 
imposing the condition $m_j=m_b$ for $j>N_{max}$. 
In order to sample the solution space, we use sharp kink profiles 
($m_j=+1$ ($-1$) for $\Delta H_1>0$ ($\Delta H_1<0$), respectively, 
if $j\le j_0$, and $m_j=m_b$ for $j> j_0$) as initial magnetization profiles. 
Starting the iteration with different initial profiles, in general,
the algorithm converges toward different local minima of the free energy. 
As already mentioned, the global free energy minimum gives the true 
equilibrium magnetization profile. Finally, we have considered different 
values of $N_{max}$, since the equilibrium profile in a complete wetting 
situation has a strong size dependence. On the contrary, no appreciable 
finite width effects occur in a partial wetting situation for $N_{max}$ 
large enough. In order to locate the wetting transition, we studied the
equilibrium profiles at bulk coexistence ($H=0$) and constant temperature, 
varying the pressure.

\begin{figure}
\includegraphics[width=8.6cm]{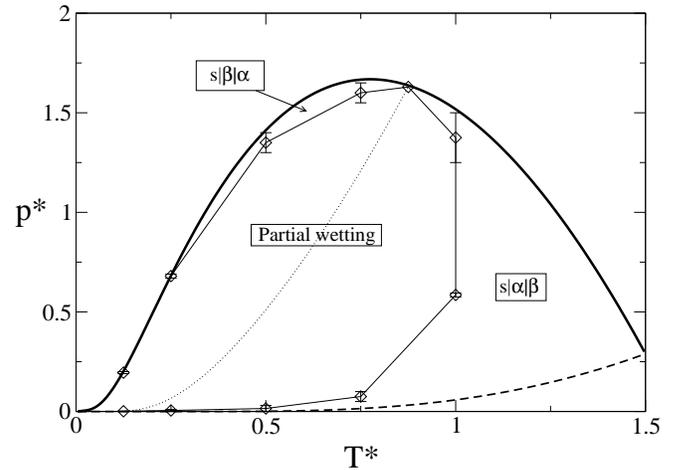}
\caption{Wetting phase diagram for the simple cubic (sc) LT model in 
the mean field approximation. The substrate-fluid interactions are 
$V_A=\epsilon_{AA}$ and $V_B=1.25\epsilon_{AA}$. The critical line 
(thick solid), A-pure vapor-liquid transition line (dashed), and 
$p_{nw}(T)$ line (dotted) are plotted. The wetting transition points (diamonds)
are also represented. The line that joins them is only for eye-guide. 
\label{fig6}}
\end{figure}

\begin{figure}
\includegraphics[width=8.6cm]{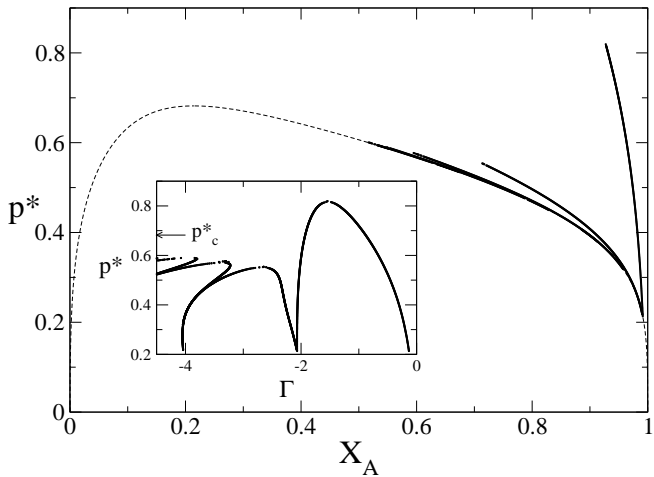}
\caption{Layering transitions for $T^*=0.25$ for the sc
LT model in the mean-field approach. The interaction couplings are the 
same as in Fig. \ref{fig6}. The dashed line correspond to the bulk 
coexistence, and the solid lines correspond to the bulk conditions 
corresponding to the layering transitions. In the inset, the coexistence 
adsorptions for the layering transition (only the first four transitions 
are plotted). The arrow shows the bulk critical pressure.  
\label{fig7}}
\end{figure}

The wetting phase diagram for $V_A=\epsilon_{AA}$ and $V_B=1.25 
\epsilon_{AA}$ is plotted in Fig. \ref{fig6}. As one can see,
the qualitative behavior of the wetting transition is similar
to the one observed in the two dimensional case. However, 
the wetting transition is associated to series of first order 
layering transitions between surface phases with the same surface 
tension, but different adsorptions. The behavior of the layering 
transitions at low temperatures shows surprising features. 
Fig. \ref{fig7} shows the first four layering transitions for $T^*\equiv 
k_B T/|\epsilon_{AA}|=0.25$. It is observed that the first layering transition
does occur for pressures higher than the bulk critical pressure. 
Consequently, for a fixed pressure between the first layering 
transition critical point and the bulk critical point corresponding 
to this temperature, the layering transition will extend to \emph{lower} 
temperatures than the critical temperature corresponding to the LCSP.   
This behavior is found only for $s|\beta|\alpha$ complete wetting in the
low temperature zone. The pressure is an increasing function
on $H$ for fixed temperature $T$ and $z_B$ (so both $K$ and $\Delta H_1$ are
fixed). Consequently, as $K>K_c$ for the layering transition critical 
point, the corresponding pressure can only be bigger than the bulk 
critical pressure if the transition occurs for $H>0$. This implies 
that $\Delta H_1$ must be negative, and thus the wetting is of type 
$s|\beta|\alpha$. The physical interpretation of this phenomenon is that 
the surface adsorbs preferentially B particles. Such a mechanism
competes with the A-B bonding (at least close to the substrate). Therefore, 
the tendency to phase separate in a fluid layer close to the substrate is 
enhanced (because the A-B bonding favors mixing) for temperatures lower than
the (lower) critical temperature. This fact was also observed in computer
simulations of a different model of associating binary mixture in Ref. 
\cite{Kumar}. However, we must stress that this phenomenon is 
not related to an enhanced coupling (i.e. the interactions between particles
in the first layer close to the substrate are increased respect the bulk 
values), that it is known to stabilize surface transitions to temperatures 
higher to the bulk critical one in the Ising model \cite{Nakanishi1}. 
On the other hand, the competition between substrate preferential 
adsorption and the A-B bonding should be also relevant for prewetting 
transitions.

\section{Confinement between symmetric walls}
 
We turn to consider the behavior of the Lin-Taylor model when confined between
two identical parallel substrates.  
Confinement in the two-dimensional case corresponds to a 
quasi-one-dimensional situation. Thus fluctuations preclude the  
existence of capillary transitions if the interactions are short-ranged.
Consequently, only the 3D case will be studied. We shall consider a
simple cubic lattice within the mean field approximation.
The substrate-fluid interactions are 
set to $V_A=\epsilon_{AA}$ and $V_B=1.25 \epsilon_{AA}$, that stabilize the 
first (surface) layering transition at temperatures lower than the lower 
critical temperature. We also assume that the separation between
plates is $N_D$ lattice units. The free energy of the confined mixture
is given by Eq. (\ref{partition}), with the mean field partition function
$Z_{Ising}(K,H,\Delta H_1)$ given by
\begin{eqnarray}
& &\lim_{N_\perp \to \infty}
\frac{1}{N_\perp}\ln Z_{Ising}(K,H,\Delta H_1)= \Delta H_1 (m_1+m_{N_D}) 
\nonumber\\
& & 
+\sum_{j=1}^{N_D}\Bigg[\frac{\nu_\parallel}{2} K m_j^2
+ \nu_\perp K m_j m_{j+1} +H m_j \nonumber\\
& &
-\left(\frac{1+m_j}{2}\right) \ln \left(\frac{1+m_j}{2}\right)
\nonumber\\
& &
- \left(\frac{1-m_j}{2}\right) \ln \left(\frac{1-m_j}{2}\right)\Bigg] 
\label{meanfield2}
\end{eqnarray}

The equation for the equilibrium profile is given by Eq. (\ref{E-L})
with the boundary condition $N_{max}=N_D/2$, for $N_D$ even, or 
$(N_D+1)/2$, for $N_D$ odd. The values of $m_j$ for $j\ge N_{max}$ are 
obtained from the symmetry of the problem: i.e. $m_j=m_{N_D-j+1}$. In 
addition to the layering transitions (related to the phenomenology 
observed on the semi-infinite case), the capillary 
transitions results from the shift of the bulk coexistence under
confinement. Near the center of the pore both phases have different
magnetizations, in contrast with the layering and prewetting transitions
where the profiles differ in a localized surface region and they have the same
values far enough from that region. In the Ising model, the shift is always 
toward higher values of $K$ (i.e. lower effective temperature) and 
$H$ negative ($H$ positive) when $\Delta H_1$ positive ($\Delta H_1$ 
negative), respectively. 

\begin{figure}
\includegraphics[width=8.6cm]{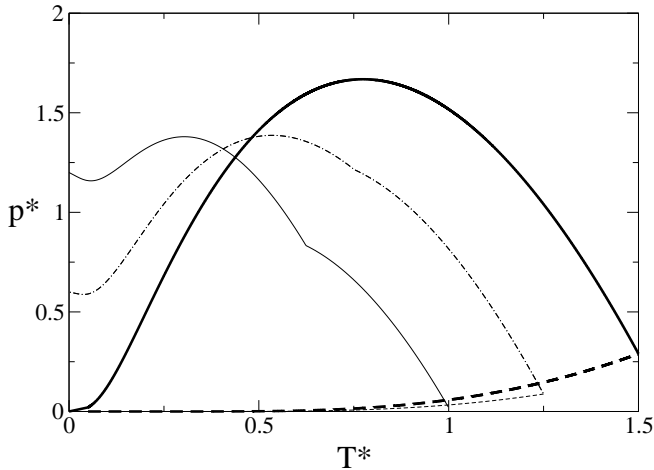}
\caption{Capillary transitions of the mean field sc LT model 
($V_A=\epsilon_{AA}$, $V_B=1.25 \epsilon_{AA}$)
for $N_D=1$ (thin solid line) and $N_D=2$ (dot-dashed line). 
In both cases, upper line is the capillary critical line, and lower line
is the A-pure capillary gas-liquid transition. The bulk conditions under 
which there is $\alpha$-$\beta$ capillary coexistence are the regions 
enclosed by their respective lines, and the $T=0$ axis. By comparison, 
the bulk critical line (thick solid line) and the A pure gas-liquid 
transition line (thick dashed line).  
\label{fig8}}
\end{figure}

The cases $N_D=1$ and $N_D=2$ can be studied analytically since $m_j=m$ for
all $j$. The bulk conditions for which the capillary transition takes place
are shown in Fig. \ref{fig8}. In spite of the fact that the temperature 
range for which the capillary transition occurs is reduced, at low 
temperatures the pressure range is largely increased. Hence, at a given 
pressure, the LCSP shifts to lower temperatures and can even disappear. 
This means that for certain values of the pressure range there are 
no immiscibility islands in the confined system, but the usual bell shaped
coexistence line terminating in a UCSP. This is a direct consequence of 
the directional character of the substrate-particle B interaction. 
The kink that it is observed in the capillary critical line corresponds to 
the crossing to the $p_{nw}(T)$ line. 

For larger values of $N_D$, the behavior of the capillary transitions are
studied numerically. As $N_D \to \infty$ the capillary critical line must 
move toward the bulk critical curve. We shall consider
the isotherm $T^*=0.25$, as a representative case 
of the low temperature regime. The excess free energy $\omega$ reads:
\begin{equation}
\omega=(\Omega + pV)/A 
\label{defomega}
\end{equation}
and converges to $2\sigma$ when $N_D\to \infty$, with $\sigma$ defined by 
Eq. (\ref{tens-LT}). The relevant order parameter for the capillary 
transitions is the average molar fraction $\bar{X}_A$, defined as
\begin{equation}
\bar{X}_A\equiv \frac{1}{N_D}\sum_{j=1}^{N_D} X_A(j) 
\label{avmol}
\end{equation}
that converges to the bulk coexistence molar fractions as $N_D \to \infty$.

\begin{figure}
\includegraphics[width=8.6cm]{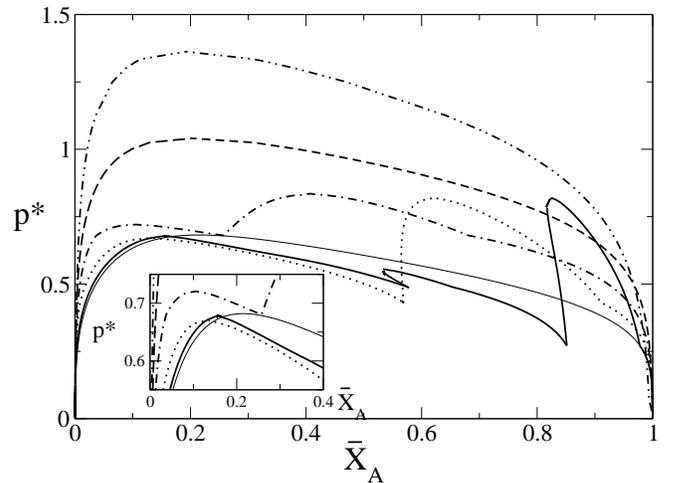}
\caption{$\bar{X}_A-p$ phase diagrams for $T^*=0.25$ and different values
of $N_D$: $N_D=1$ (dashed-double dotted line), $N_D=2$ (dashed line), 
$N_D=5$ (dotted line) and $N_D=15$ (thick solid line), under the same 
conditions as in Fig. \ref{fig8}. The bulk transition (thin solid line) 
is also plotted for comparison. The inset corresponds to an enlargement of 
the zone around the bulk critical point. \label{fig9}}
\end{figure}

We have obtained the constant temperature phase diagrams for $T^*=0.25$ (see 
Fig. \ref{fig9}). Various transitions are observed for $N_D\ge 3$. One of
these transitions converges toward the bulk phase transition for large 
separations and is identified with the true capillary transition. The 
remaining transitions are characterized by having similar average molar 
fractions and correspond to the layering transitions observed in the 
presence of a single substrate. The coverages (as defined by Eq. 
(\ref{adsorcion})) corresponding to the coexisting phases in these 
transitions converge very fast to the semi-infinite values, as expected 
from their surface nature. However, the number of these transitions 
depends strongly on $N_D$, since the capillary transition, which 
depends sensibly on $N_D$, competes with them \cite{Bruno,Tarazona2}. 
For instance, in the case of $3\le N_D <10$ only the first layering 
transition appears, while the second appears for $N_D\ge 10$. 

\begin{figure}
\includegraphics[width=8.6cm]{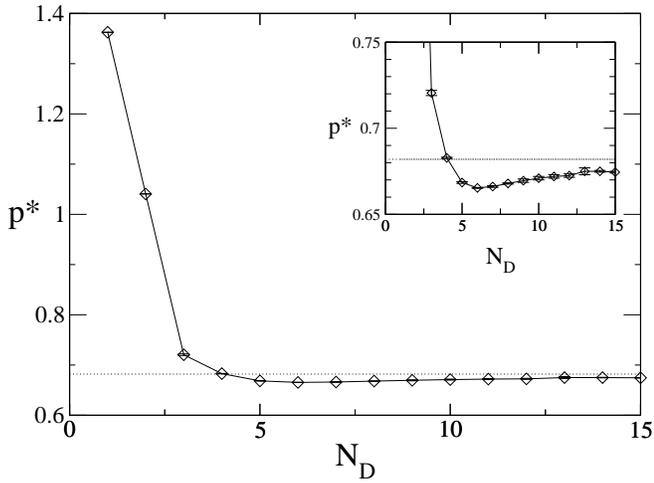}
\caption{Capillary reduced critical pressures $p^*$ versus $N_D$ for 
$T^*=0.25$ (same conditions as Fig. \ref{fig8}). The dotted line corresponds
to the bulk critical pressure. Inset: enlargement of the minimum zone. 
\label{fig10}}
\end{figure}

The values of the capillary critical pressure for small $N_D$ are larger
than the corresponding bulk critical values. By increasing $N_D$, the capillary 
critical pressure decreases up to a minimum, corresponding to a pressure less
than the bulk critical pressure. From such value it converges
monotonically to the bulk critical value (see figure \ref{fig10}).
This fact means that along an isobaric the LCSP shifts to lower temperatures
only for very narrow pores. Consequently, the stabilizing mechanism of 
the coexistence region below the bulk LCSP seems to be only effective for 
small values of $N_D$. The deviations of the capillary critical pressure with 
respect to the bulk value in the range $5 < N_D \le 15$ follow a power law
$\Delta p \sim N_D^{-x}$, with $x = 1.04 \pm 0.09$. This behavior is consistent
with the findings for the confined lattice gas for intermediate values of $N_D$
reported in Ref. \cite{Bruno}. The scaling behavior $\Delta p \sim 
N_D^{-1/\nu}$ ($N_D^{-2}$ in the mean-field approach) predicted by the 
Nakanishi-Fisher theory should be obeyed for larger values of $N_D$. However, 
our results do not extend to that region due to the numerical error in 
estimating the capillary critical pressure close to the bulk 
critical point. Nevertheless, we do not expect a change in the tendency for 
large values of $N_D$ and we expect our results to be qualitatively 
correct for all values of $N_D$. It is interesting to note that the
slab approximation \cite{Bruno}, which corresponds to set $m_j=m$ for all $j$,
predicts that for sufficiently small $T^*$ the capillary critical pressure
is larger than the bulk value for \emph{all} the values of $N_D$.
The failure of this approach is not surprising since it gives poor results
for the confined lattice gas close to the capillary critical point 
as it overestimates both $K$ and $H$ at the capillary critical 
point \cite{Bruno}. This fact alters the balance between the different 
mechanisms involved in the shift of the critical point, leading to the 
wrong result mentioned above. 

\section{The cylindrical pore network}

\begin{figure}
\includegraphics[width=8.6cm]{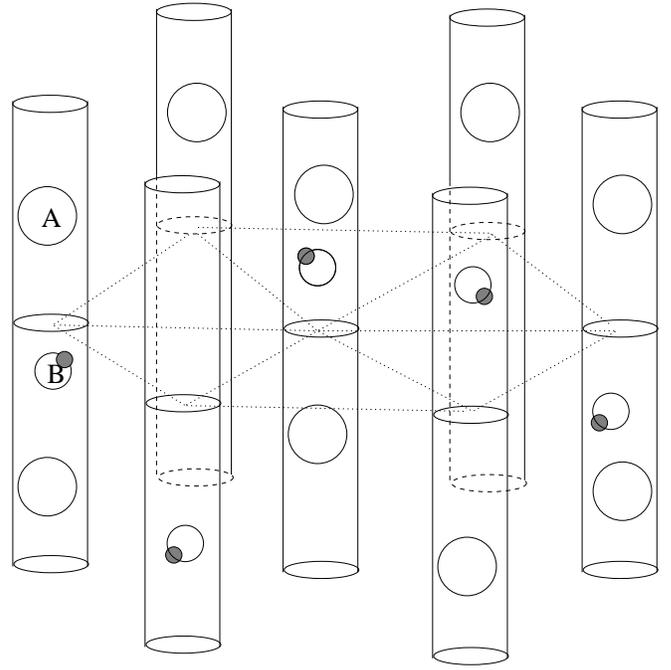}
\caption{Schematic representation of the LT model adsorbed in a network of
narrow infinitely long cylinders. The A particles are represented as spherical
particles, and the B particles as spherical particles with an interaction 
side (shadow circle). In this case, the coordination number is 
$\nu_\perp=6$. 
\label{fig11}}
\end{figure}

Finally, we consider the confinement of the LT mixture in a 
network of parallel infinitely long, cylindrical pores. This model
can be understood as a crude approximation for adsorption in zeolites. 
It represents another instance in which the substrate can
prevent the formation of hydrogen bonds: the geometrical hindrance due
to the confinement in very narrow nanopores. We shall assume that the 
pore diameter is so small that the A-B bonds can form 
only along the direction $z$ parallel to the cylinder axis (see figure 
\ref{fig11}). The cylinder diameter is taken to be 
less than twice the A-B collision diameter. In addition, the pores have
diameters less than two A particle diameters and they form a two 
dimensional network with coordination number $\nu_\perp$. We allow 
attenuated interactions between particles in nearest neighbor pores, 
in order to allow a capillary transition. A similar zeolite model has been 
employed to study the capillary transition of methane adsorbed in 
$\textrm{ALPO}_4$-5 \cite{Radhakrishnan}. 

The in-pore fluid-fluid interactions have the same form as in bulk, with 
coupling constants $\epsilon_{AA}$, $\epsilon_{AB}$ and $\epsilon_{BB}$. 
However, the interactions between particles in nearest-neighbor pores are
attenuated to $\tilde{\epsilon}_{AA}$, $\tilde{\epsilon}_{AB}$ and 
$\tilde{\epsilon}_{BB}$ ($|\tilde{\epsilon}_{ij}| < |\epsilon_{ij}|$), 
although the A-B and B-B interactions directional character remains the 
same as before. This system behaves as a bulk lattice model with anisotropic 
interactions. The grand-canonical free energy per unit volume reads:
\begin{eqnarray}
\frac{\Omega}{V}&=&
-\frac{k_B T}{v_0}\Bigg[\frac{\nu_\perp}{2} 
\ln\left(1+2\tilde{z}_B
+\tilde{z}_B^2\mathrm{e}^{-\beta \tilde{\epsilon}_{BB}}\right)\nonumber\\
&+&\ln(1+2z_B+z_B^2 \mathrm{e}^{-\beta \epsilon_{BB}}) + H - \frac{\nu_\perp 
K_\perp}{2} \nonumber \\ &-&K_\parallel+ \frac{1}{N} \ln Z_{Ising} 
(K_\perp,K_\parallel,H)\Bigg]
\label{free-energy-zeolite}
\end{eqnarray}
where $\tilde{z}_B \equiv z_B {\rm e}^{-\beta V_B}$ is the effective activity
of the $B$ particles oriented in the plane perpendicular to the cylinder axis,
and $Z_{Ising}$ is canonical partition function of an anisotropic 
Ising model characterized by couplings $K_\parallel$ in the parallel 
directions to the cylinder axis and $K_\perp$ in the perpendicular directions,
and an effective magnetic field $H$. $K_\parallel$ has the same expression
as the bulk coupling constant Eq. (\ref{defk}), while $K_\perp$ and $H$ 
are defined as:
\begin{eqnarray}
K_{\perp}&=&-\frac{\beta \tilde{\epsilon}_{AA}}{4}
+\frac{1}{2} \ln\left(\frac{\sqrt{1+2\tilde{z}_B+ 
\tilde{z}_B^2 \mathrm{e}^{-\beta  
\tilde{\epsilon}_{BB}}}}{1+\tilde{z}_B \mathrm{e}^{-\beta 
\tilde{\epsilon}_{AB}}}\right)
\label{defK2-zeolite}\\
\nonumber\\
\tilde{H}&=&H+\nu_\perp \Bigg[\frac{\beta}{4} \left(\epsilon_{AA}-
\tilde{\epsilon}_{AA}-\frac{2V_A}{\nu_\perp}\right)\nonumber\\
&+& \frac{1}{4} \ln \left(\frac{1+2z_B+z_B^2 \mathrm{e}^{-\beta 
\epsilon_{BB}}}{1+2\tilde{z}_B +
\tilde{z}_B^2\mathrm{e}^{-\beta \tilde{\epsilon}_{BB}}} 
\right)\Bigg] \label{defH-zeolite}
\end{eqnarray}
with $H$ given by Eq. (\ref{defh}). One can see that confinement shifts 
the effective magnetic field $H$ with respect to its bulk value 
with a contribution analogous to that found for planar substrates
(compare Eqs. (\ref{H_1}) and (\ref{defH-zeolite})). In addition, the 
confinement determines an anisotropy in the couplings.
The A particle molar fraction $X_A$ reads:
\begin{equation}
X_A=\frac{n_A}{n_A+n_B^1+n_B^2}
\label{fracmolzeo}
\end{equation}
where $n_A=(1+m)/2v_0$ is the A particle density, and $n_B^1$ ($n_B^2$) is
the density of B particles oriented in a perpendicular (parallel) direction 
to the cylinder axis, respectively:
\begin{eqnarray}
n_B^1 &=& \frac{\nu_\perp}{4 v_0} \left(\frac{\tilde{z}_B + \tilde{z}_B^2 
\mathrm{e}^{-\beta\tilde{\epsilon}_{BB}}}
{ 1 + 2 \tilde{z}_B + \tilde{z}_B^2 \mathrm{e}^{-\beta
\tilde{\epsilon}_{BB}}} + \frac{\tilde{z}_B \mathrm{e}^{-\beta 
\tilde{\epsilon}_{AB}}}{1 + \tilde{z}_B \mathrm{e}^{-\beta 
\tilde{\epsilon}_{AB}}}\right)\nonumber \\
&-& \frac{\nu_\perp}{4v_0} \left(\frac{\tilde{z}_B + \tilde{z}_B^2 
\mathrm{e}^{-\beta\tilde{\epsilon}_{BB}}}
{1 + 2 \tilde{z}_B + \tilde{z}_B^2 \mathrm{e}^{-\beta
\tilde{\epsilon}_{BB}}} -\frac{\tilde{z}_B 
\mathrm{e}^{-\beta\tilde{\epsilon}_{AB}}}
{1 + \tilde{z}_B \mathrm{e}^{-\beta 
\tilde{\epsilon}_{AB}}}\right)\nonumber \\
&\times&\left(\frac{\partial \ln Z_{Ising}/N}{\partial
K_{\perp}}\right)\nonumber\\
&-& \frac{\nu_\perp}{2v_0} \frac{\tilde{z}_B + \tilde{z}_B^2 
\mathrm{e}^{-\beta \tilde{\epsilon}_{BB}}}{1 + 2 \tilde{z}_B + 
\tilde{z}_B^2 {\rm e}^{-\beta \tilde{\epsilon}_{BB}}}m 
\label{densb1-zeolita}\\
\nonumber\\
n_B^2 &=& \frac{1}{2 v_0} \left(\frac{z_B + z_B^2 \mathrm{e}^{-\beta 
\epsilon_{BB}}}{1 + 2 z_B + z_B^2 \mathrm{e}^{-\beta \epsilon_{BB}}} +
\frac{z_B \mathrm{e}^{-\beta \epsilon_{AB}}}{1 + z_B \mathrm{e}^{-\beta
\epsilon_{AB}}}\right)\nonumber \\
&-& \frac{1}{2v_0} \left(\frac{z_B + z_B^2 \mathrm{e}^{-\beta \epsilon_{BB}}}
{1 + 2 z_B + z_B^2 \mathrm{e}^{-\beta \epsilon_{BB}}} -
\frac{z_B \mathrm{e}^{-\beta \epsilon_{AB}}}{1 + z_B \mathrm{e}^{-\beta
\epsilon_{AB}}}\right)\nonumber \\
&\times&\left(\frac{\partial \ln Z_{Ising}/N}{\partial
K_{\parallel}}\right)\nonumber\\
&-& \frac{1}{v_0} \frac{z_B + z_B^2 \mathrm{e}
^{-\beta \epsilon_{BB}}}{1 + 2 z_B + z_B^2 \mathrm{e}^{-\beta 
\epsilon_{BB}}}m 
\label{densb2-zeolita}
\end{eqnarray}

Phases characterized by different molar fractions $X_A$ may coexist when the 
spontaneous (i.e. $\tilde{H}=0$) magnetization $m=\pm|m|\neq 0$. We 
consider the case $\epsilon_{AA}=-1$, $\epsilon_{AB}=-0.6$ and 
$\epsilon_{BB}=0$, that corresponds to a case that presents closed-loops 
of immiscibility in the bulk. On the other hand, as the hydrogen bond 
interactions are very short-ranged, we will take $\tilde{\epsilon}_{AB}=
\tilde{\epsilon}_{BB}=0$ as physically sensible values. Moreover, 
$\tilde{\epsilon}_{AA}$ is set to be $\alpha \epsilon_{AA}$, where 
$\alpha$ is a positive number less than one. 

In order to discriminate between different mechanisms, we chose the 
substrate-fluid interactions in such a way that $\tilde{H}=H$, i.e. 
$V_A=\nu_\perp (1-\alpha)\epsilon_{AA}/2$ and $V_B=0$. Hence, the 
preferential substrate adsorption effect (very similar to the one studied 
in the slab geometry) is suppressed and we can focus on the coupling 
anisotropy. 

We have studied the 2D square lattice case, for which analytical expressions
for $Z_{D,\nu}$ are known in the anisotropic case \cite{Onsager}. However,
we expect that the features obtained for this case will be qualitatively 
correct also for the 3D case. From the exact solution, the values of 
$(K_\perp,K_\parallel)$ that correspond to the phase coexistence satisfy 
the following relationship:
\begin{equation}
\sinh(2K_\parallel) \sinh(2K_\perp)\ge 1
\label{coex-Ising-anis}
\end{equation}
where equality holds only at the critical point. It is clear from such an 
expression that, even for $K_\parallel$ (equal to the bulk value) 
less than the isotropic bulk value, the system can undergo a 
phase transition for $K_\perp$ large enough. Note that since $K_\perp=-\beta 
\alpha \epsilon_{AA}/4$, this condition is fulfilled in the low 
temperature region. This fact is also observed in the
mean-field approach, for which the coexistence condition is 
$\nu_\perp K_\perp + 2 K_\parallel \ge 1$. 
 
\begin{figure}
\includegraphics[width=8.6cm]{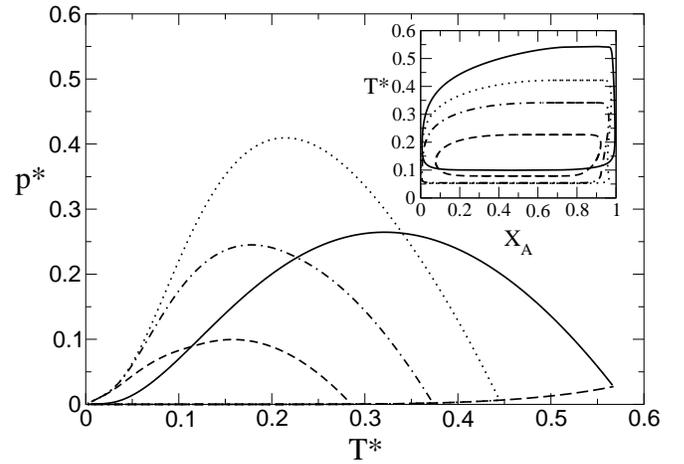}
\caption{Capillary transitions in the network of cylindrical pores. The 
reservoir conditions corresponding to the capillary critical lines are
represented for different values of $\alpha$: $\alpha=0.2$ (dashed line), 
$\alpha=0.4$ (dotted-dashed line) and $\alpha=0.6$ (dotted line). 
The bulk critical line (solid line) and the A pure vapor-liquid line 
(long dashed line) are also plotted. In the inset, the phase diagrams 
corresponding to $p^*=0.07$ for the same values of $\alpha$ (the symbol 
meaning is the same as previously) and their comparison with the bulk 
phase diagram (solid line).
\label{fig12}}
\end{figure}

Figure \ref{fig12} shows the reservoir conditions under which the capillary
phase transition occurs for different values of $\alpha$. There also exist 
immiscibility islands for the capillary transition within the porous 
medium (see inset in Fig. \ref{fig12}), but the corresponding LCSP occurs 
at temperatures lower than those of the bulk LCSP at the same pressure. 
As $\alpha$, increases the temperature and pressure range for which one 
observes capillary transition increases. Hence, the immiscibility islands 
become larger with $\alpha$, but mainly in the high temperature region. 
In fact, the temperature corresponding to the LCSP is rather insensitive 
to the value of $\alpha$, and it is always below the temperature 
corresponding to the bulk LCSP. Our findings can be understood by a simple
physical argument. The geometry forbids the formation of hydrogen bondings 
in the directions perpendicular to the cylinder axis. So, the confinement 
effectively reduces the A-B bonds respect to the bulk case, and consequently 
the phase separation is enhanced in the lower temperature regime, in complete
agreement with our results.

\section{Discussion and conclusions}

In this paper we have studied the effect on the phase equilibria of an 
associating binary mixture in the presence of substrates and under 
confinement in very simple geometries. We have focused on the conditions 
in which bulk system exhibits closed-loops of immiscibility. Our results 
show clearly that the presence of directional fluid-substrate interactions 
can stabilize purely surface transitions (such as layering and prewetting) 
and also for the phase separation under confinement in nanopores, for 
temperatures below the bulk LCSP. However, this mechanism becomes 
inefficient for the capillary transition when the surface-to-volume ratio 
is small, and consequently we doubt that it can explain the experimental 
results in Refs. \cite{Scheibner,Meadows}, which would correspond to 
the latter case. 
The disagreement with the scaling predictions in the cases in which the
temperature shift depends on the substrate remains, in any case, unsolved.
The experimental data show a capillary critical temperature dependence on
the film width that it is not consistent with the scaling predictions
\cite{Nakanishi3,Nakanishi4} 
and it could be a signature of an emerging new characteristic length scale.
However, to our knowledgement there is no theoretical explanation of such
a behavior.

Some final remarks are pertinent. Firstly, the present model 
represents a simplified description of an associating binary mixture, but 
keeps all the physical ingredients needed to reproduce the correct 
phenomenology. Consequently, we are confident that our conclusions can be 
relevant also for more sophisticated models. Secondly, fluid adsorption 
in nanotubes and zeolites is experimentally feasible and they could 
provide experimental probes of our theoretical predictions.

\acknowledgments

The authors wish to thank Dr. I. Rodr\'{\i}guez-Ponce, Prof. G. Jackson 
and Prof. A. O. Parry for their interest and useful comments on this paper.
J.M.R.-E. and L.F.R. gratefully acknowledge financial support for this 
research by Grant No. PB97-0712 from DGICyT (Spain) and No. FQM-205 
from PAI (Junta de Andaluc\'{\i}a). J.M.R.-E. also wishes to thank Ministerio
de Educaci\'on, Cultura y Deporte (Spain) for partial financial support. 
Finally U.M.B.M acknowledges financial support by Ministero dell'Istruzione, 
dell'Universit\'a e della Ricerca, Cofin 2001 Prot. 2001023848.

\end{document}